\documentclass[preprint2]{aastex}
\usepackage{color}
\usepackage{epsfig}
\usepackage[fleqn]{amsmath}
\usepackage{multirow}
\usepackage{multicol}
\usepackage{mathptmx}
\usepackage{graphicx}
\usepackage{bm}
\usepackage{mathrsfs}
\usepackage{lineno}

\shorttitle{Chaos} \shortauthors{Yang et al.}
\begin{document}
\title{\large Chaos in a Magnetized  Modified Gravity Schwarzschild Spacetime}
\author{Daqi Yang$^{1}$, Wenfu Cao$^{1}$, Naying Zhou$^{1}$, Hongxing Zhang$^{1}$, Wenfang Liu$^{1}$, Xin Wu$^{1,2 \dag}$}
\affil{ 1. School of Mathematics, Physics and Statistics $\&$
Center of Application and Research of Computational Physics,
Shanghai
University of Engineering Science, Shanghai 201620, China; \\
m440121502@sues.edu.cn (D.Y.); m440121503@sues.edu.cn (W.C.); m130120101@sues.edu.cn (N.Z.);
m130120111@sues.edu.cn (H.Z.); 21200007@sues.edu.cn (W.L.) \\
2. Guangxi Key Laboratory for Relativistic Astrophysics, Guangxi
University, Nanning 530004, China} \email{$\dag$ Corresponding
Author: xinwu@gxu.edu.cn or wuxin$\_$1134@sina.com. }

\begin{abstract}

Based on the scalar-tensor-vector modified
gravitational theory, a modified gravity Schwarzschild black hole
solution has been given in the existing literature. Such a black
hole spacetime is obtained through the inclusion of a modified
gravity coupling parameter, which corresponds to the modified
gravitational constant and the black hole charge. In this sense,
the modified gravity parameter acts as not only an enhanced
gravitational effect but also a gravitational repulsive force
contribution to a test particle moving around the black hole.
Because the modified Schwarzschild spacetime is static spherical
symmetric, it is integrable. However, the spherical symmetry and
the integrability are destroyed when the black hole is immersed in
an external asymptotic uniform magnetic field and the particle is
charged. Although the magnetized modified Schwarzschild spacetime
is nonintegrable and inseparable, it allows for the application of
explicit symplectic integrators when its Hamiltonian is split into
five explicitly integrable parts. Taking one of the proposed
explicit symplectic integrators and the techniques of Poincar\'{e}
sections and fast Lyapunov indicators as numerical tools, we show
that the charged particle can do chaotic motions under some
circumstances. Chaos is strengthened with an increase of the
modified gravity parameter from the global phase space structures.
There are similar results when the magnetic field parameter and
the particle energy increase. However, an increase of the particle
angular momentum weakens the strength of chaos.

\end{abstract}

\section{Introduction}

The Einstein field equations in general relativity predict the
existence of black hole solutions, such as the Schwarzschild,
Reissner-Nordstr\"{o}m and Kerr metrics. In recent several years,
a number of detections of gravitational-wave signals emitted from
binary black hole mergers [1,2] and the  Event Horizon Telescope
(EHT) shadow image of M87$^{*}$ central supermassive black hole
[3] have frequently confirmed the prediction.

Although the great success of General Relativity
has been achieved, developments in constructing alternative
theories of gravity are necessary due to the requirement of rapid
progress in the field of observational cosmology [4]. In fact,
several notable examples like Eddington's theory of connections,
Weyl's scale independent theory and the higher dimensional
theories of Kaluza and Klein were made during the very early days
after Einstein's theory of General Relativity. The Eddington's
theory shows the consistency of the magnitude of a varying
Newton's constant and the ratio of the mass and scale of the
Universe. The Parameterised Post-Newtonian (PPN) formalism by
Kenneth Nordtvedt, Kip Thorne and Clifford Will allows for
precision tests of fundamental physics on the scale of the
observable Universe. The limits of General Relativity occur for
the emergence of the dark universe scenario. The dark energy
beyond Einstein's theory is good to explain the apparent
accelerating expansion of the Universe. This shows that General
Relativity may not be suitable for describing the Universe on the
largest scales. Constructing a quantum field theory of gravity is
based on the rise of super-gravity and super-string theories. The
black hole singularity problem in the general relativistic black
hole solutions should be avoided in the study of some theories of
gravity like quantum field theory [5-9]. In short, many
experimental tests and theoretical studies of strong-field gravity
features often require that such a strong gravitational field
should not be one described by the standard general relativity but
should be a departure from the general relativity. 

The review on ``Modified gravity and cosmology"
[4] provides a useful reference tool for researchers and students
in cosmology and gravitational physics. Many modified gravity
theories with extra fields to the Einstein's theory of general
gravity were introduced in [4]. Some examples are
quantum-corrected gravity theories [10-15] including Kaluza-Klein
gravity theories [16,17], scalar-tensor theories [18-23],
Einstein-$\AE$ther theories [24], Bimetric theories [25], $f(R)$
theories [26,27], $f(T)$ gravity [28], and scalar-tensor-vector
gravity [29,30]. The quantum-corrected gravity theories relate to
higher dimensional gravity theories including extra spatial
dimensions and extra temporal dimensions. The Kaluza-Klein theory
is devoted to unifying gravity and electrodynamics, and its basic
idea is based on General Relativity built on a 4+1 dimensional
manifold with one small and compact spatial dimension. The
scalar-tensor theories of gravity are established through the
Lagrangian density with the metric tensor coupling to scalar field
and matter fields. They allow possible variations in Newton's
constant, $G_N$. The $f(R)$ theories of gravity are derived from a
generalisation of the Einstein-Hilbert density. They are useful to
explain the observed accelerating expansion of the Universe. The
scalar-tensor-vector gravity theory contains a vector field, three
scalar fields and a metric tensor. It can well explain the solar
observations, the rotation curves of galaxies [31]  and the
dynamics of galactic clusters [32]. Based on the theory of
gravity, a static spherically symmetric modified gravity
Schwarzschild black hole metric was first given in [33]. The
metric describes the final stage of the collapse of a body by
introducing $\alpha$ as a coupling parameter of modified gravity,
which enhances the gravitational constant and provides a charge
yielding a gravitational repulsive force. In fact, the modified
gravity Schwarzschild black hole \emph{seems} to be a
Reissner-Nordstr\"{o}m black hole by the modified gravity coupling
parameter adjusting the gravitational constant and acting as the
black hole charge. 

The authors of [34] studied circular orbits of charged particles
around the modified gravity Schwarzschild black hole immersed in
an asymptotically uniform magnetic field. They found that no
stable circular orbits exist when the magnetic coupling parameter
is not smaller than 1. The range of stable circular orbits
increases as the modified gravity  coupling parameter and the
magnetic coupling parameter increase. The center-of-mass energy
collision of charged particles increases with  the modified
gravity coupling parameter increasing. The authors of [35] also
showed that the innermost stable circular orbits and marginally
bound orbits in the modified gravity Schwarzschild metric are
larger than those in the pure Schwarzschild spacetime. The
positions of the innermost stable circular orbits for charged
particle are less than those for neutral particles. In addition,
the shadow cast by the spherical symmetric black hole in the
modified gravity was investigated. When the modified gravity
coupling parameter increases, the sizes of photonsperes and
shadows of the black hole are enlarged and can be observed through
EHT [36].

The authors of [34,35] mainly surveyed the effect of the modified
gravity  coupling parameter on the circular motions of charged
particles at the equatorial plane. Unlike them, we shall consider
the modified gravity  coupling parameter how to effect the regular
and chaotic orbital dynamics of charged particles in the global
phase space. For the sake of our purpose, a dynamical model for
the description of charged particles moving near the modified
gravity Schwarzschild black hole with an external magnetic field
is introduced in Section 2. Then, explicit symplectic methods are
designed for this dynamical problem and the orbital dynamics of
charged particles is explored in Section 3. Finally, our main
results are concluded in Section 4.

\section{Modified Gravity Nonrotating Black Hole Immersed in an External Magnetic Field}

In terms of the scalar-tensor-vector modified gravitational
theory, a static spherically symmetric nonrotating  black hole
[33] is written in Boyer-Lindquist coordinates
$x^{\mu}=(t,r,\theta ,\varphi)$ as
\begin{eqnarray}
ds^2 &=& g_{\mu \nu}dx^\mu dx^\nu \nonumber \\
&=& -fc^2dt^2+\frac{1}{f}dr^2 \nonumber \\ && +r^{2}(d\theta
^2+\sin^2\theta d\varphi ^2),
\end{eqnarray}
where function $f$ has the following form
\begin{eqnarray}
    f = 1-\frac{2(1+\alpha)G_NM}{rc^2}+\frac{\alpha(1+\alpha)G^2_NM^2}{r^2c^4}.
\end{eqnarray}
Several notations are specified here. $c$ is the speed of light
and $G_N$ represents the Newton's gravitational constant. $M$
stands for the black hole mass. $\alpha$ is a
dimensionless modified gravity  coupling parameter, which is
responsible for adjusting the gravitational constant $G_N$ as
$G=G_N(1+\alpha)$ and providing the black hole charge $Q=\pm
M\sqrt{\alpha G_N}$. For $\alpha>0$, the adjusted gravitational
constant $G$ is larger than the the Newton's gravitational
constant $G_N$; this implies that $\alpha$ in the second term of
Equation (2) can enhance the gravitational effects. However,
$\alpha$ in the third term of Equation (2) gives the black hole
charge with a gravitational repulsive force. Thus, the modified
gravity parameter plays roles in inducing an enhanced
gravitational effect and a gravitational repulsive force
contribution. In other words, Eq. (1) for the description of the
modified gravity Schwarzschild metric \emph{looks like}  the
Reissner-Nordstr\"{o}m black hole metric when $G_N(1+\alpha)$ and
$\pm M\sqrt{\alpha G_N}$ in Eq. (2) are respectively replaced by
the adjusted gravitational constant $G$ and the charge $Q$,
$G_N(1+\alpha)\rightarrow G$ and $\pm M\sqrt{\alpha
G_N}\rightarrow Q$.  There are two horizons
$r_{\pm}=G_NM(1+\alpha\pm\sqrt{1+\alpha})/c^2$ for $\alpha>0$.
$\alpha=0$ corresponds to the Schwarzschild event horizon
$r_{+}=r_S=2G_NM/c^2$.

Assume that the black hole is immersed in an asymptotically
uniform external magnetic field, whose four-vector potential
satisfying the Maxwell equation in the curved spacetime background
has a nonzero component [34,35]
\begin{eqnarray}
    A_\varphi =\frac{1}{2}B[r^2-\alpha(1+\alpha)M^2]\sin^2\theta.
\end{eqnarray}
Parameter $B$ is the magnetic field strength.

Consider that a particle with mass $m$ and charge $q$ moves around
the modified gravity Schwarzschild black hole surrounded by the
external magnetic field. The particle motion is described in the
following Lagrangian
\begin{eqnarray}
\mathcal{L}  =\frac{m}{2}g_{\mu \nu}\dot{x}^\mu
\dot{x}^\nu+qA_\mu\dot{x}^\mu,
\end{eqnarray}
where $\dot{x}^\mu$ is the 4-velocity, i.e., the derivative of
coordinate $x^{\mu}$ with respect to the proper time $\tau$. The
covariant generalized 4-momentum is defined by
\begin{eqnarray}
    p_\mu =\frac{\partial \mathcal{L} }{\partial \dot{x}^\mu }=mg_{\mu \nu}
\dot{x}^\nu+qA_\mu.
\end{eqnarray}
Based on the Euler-Lagrangian equations, two components of the
4-momentum are conserved, that is,
\begin{eqnarray}
    p_t &=& -mf\dot{t}=-E, \\
    p_\varphi &=& mr^2\sin^2 \theta \dot{\varphi }+q A_\varphi=L.
\end{eqnarray}
$E$ is the energy of the particle, and $L$ denotes the angular
momentum of the particle. This Lagrangian is equivalent to the
Hamiltonian
\begin{eqnarray}
    H=\frac{1}{2m}g^{\mu \nu}(p_\mu -qA_\mu)(p_\nu-qA_\nu).
\end{eqnarray}

For simplicity, $c$ and $G_N$ are taken as geometric units:
$c=G_N=1$. In addition, dimensionless operations are implemented
through scale transformations: $r \to rM$, $t \to Mt$, $\tau \to
M\tau$, $B \to B/M$, $E \to mE$, $L \to mML$, $p_r \to mp_r$,
$p_\theta\to mMp_\theta$, $q \to mq$, $H \to mH$. Thus, $m$ and
$M$ in the above-mentioned expressions are also used as geometric
units $m = M = 1$. Now, the Hamiltonian has a simple expression
\begin{eqnarray}
H &=& \frac{p_r^2}{2}[1-\frac{2(1+\alpha)}{r}
+\frac{\alpha(1+\alpha)}{r^2}]+\frac{1}{2}\frac{p_\theta^2}{r^2} \nonumber \\
&& + \frac{1}{8r^2}[\frac{2L}{\sin\theta}+\beta (\alpha^2+\alpha-r^2)\sin\theta]^2 \nonumber \\
    && -\frac{E^2r^2}{2[\alpha+\alpha^2-2r(1+\alpha)+r^2]},
\end{eqnarray}
where $\beta =Bq$. This system has two degrees of freedom in a
four-dimensional phase space made of $(r,\theta,p_r,p_\theta)$.

If the spacetime (1) is time-like, the Hamiltonian is always
identical to a given constant
\begin{eqnarray}
    H=-\frac{1}{2}.
\end{eqnarray}
The system (9) is inseparable to the variables. In this case, no
other constants of motion but the three constants in Eqs. (6), (7)
and (10) are present. Numerical integration methods are convenient
to solve such a nonintegrable system.

\section{Numerical Investigations}

Several explicit symplectic integrators are designed for the
Hamiltonian (9). Then, one of the algorithms is used to provide
some insight into the regular and chaotic dynamics of charged
particle orbits in the system (9).

\subsection{Construction of Explicit Symplectic Methods}

It is clear that the Hamiltonian (9) is not split into two parts
with analytical solutions as explicit functions of proper time and
then does not allow for the application of explicit symplectic
algorithms. However, the explicit symplectic methods are still
available when the Hamiltonian describing the motion of a charged
particle around the Reissner-Nordstr\"{o}m black hole immersed in
an external magnetic field is separated into five parts having
explicitly analytical solutions, as was claimed in [37]. The idea
on the construction of explicit symplectic integrators is also
applicable to the Hamiltonian (9) for the description of the
motions of charged particles around the modified gravity
Schwarzschild black hole. The related details are given to the
algorithmic construction.

Following the work [37], we split the Hamiltonian (9)  into five
parts
\begin{eqnarray}
    H=H_1+H_2+H_3+H_4+H_5,
\end{eqnarray}
where all sub-Hamiltonians are expressed as
\begin{eqnarray}
    H_1 &=& \frac{1}{8r^2}[\frac{2L}{\sin\theta}+\beta (\alpha^2+\alpha-r^2)\sin\theta]^2 \nonumber \\
    && -\frac{E^2r^2}{2[\alpha+\alpha^2-2r(1+\alpha)+r^2]},\\
    H_2 &=& \frac{1}{2}p_r^2,\\
    H_3 &=& -\frac{(1+\alpha)}{r}p_r^2,\\
    H_4 &=& \frac{p_\theta  ^2}{2r^2},\\
    H_5 &=& \frac{1}{2}\frac{\alpha(1+\alpha)}{r^2}p_r^2.
\end{eqnarray}
The sub-Hamiltonians $H_2$ and $H_4$ are consistent with those in
[37], but the others are somewhat different. The five splitting
parts are solved analytically and their analytical solutions are
explicit functions of proper time $\tau$. Operators for
analytically solving these sub-Hamiltonians are $\mathcal{H}_1$,
$\mathcal{H}_2$, $\mathcal{H}_3$, $\mathcal{H}_4$ and
$\mathcal{H}_5$. The splitting method is based on
the case of $\alpha>0$. If $\alpha=0$, then $H_5=0$ and the
Hamiltonian (9) has four explicitly integrable splitting parts.
This case is the same as the Reissner-Nordstr\"{o}m black hole
with a vanishing charge in [37].

Setting $h$ as a proper time step, we define two first-order
operators
\begin{eqnarray}
\aleph (h)&=& \mathcal{H}_1(h)\times\mathcal{H}_2(h)\times\mathcal{H}_3(h)\nonumber\\
&&\times\mathcal{H}_4(h)\times\mathcal{H}_5(h), \\
\aleph^{\ast} (h) &=&
\mathcal{H}_5(h)\times\mathcal{H}_4(h)\times\mathcal{H}_3(h)\nonumber\\
&&\times\mathcal{H}_2(h)\times\mathcal{H}_1(h).
\end{eqnarray}
The product of $\aleph^{\ast}$ and $\aleph$ is a symmetric
composition as an explicit symplectic algorithm to a second-order
accuracy
\begin{eqnarray}
S_2(h)=\aleph^{\ast}(\frac{h}{2})\times\aleph(\frac{h}{2}).
\end{eqnarray}
This method can rise to a fourth-order accuracy [38]
\begin{eqnarray}
    S_4(h)=S_2(\gamma h)\times S_2(\delta h)\times S_2(\gamma h),
\end{eqnarray}
where $ \gamma =1/(1-\sqrt[3]{2}) $ and $ \delta =1-2\gamma$.
There is an optimized fourth-order partitioned Runge-Kutta (PRK)
explicit symplectic integrator [39]
\begin{eqnarray}\nonumber
PRK_64(h) &=& \aleph^{\ast} (\alpha _{12} h)\times \aleph(\alpha
_{11}
h)\times \cdots \\
&& \times \aleph^{\ast} (\alpha _2 h) \times \aleph(\alpha _1 h),
\end{eqnarray}
where time-step coefficients are listed in [40] by
\begin{eqnarray}
    \nonumber
    &&\alpha _1=\alpha _{12}= 0.0792036964311597,   \\ \nonumber
    &&\alpha _2=\alpha _{11}= 0.1303114101821663,   \\ \nonumber
    &&\alpha _3=\alpha _{10}= 0.2228614958676077,    \\ \nonumber
    &&\alpha _4=\alpha _9=-0.3667132690474257,    \\ \nonumber
    &&\alpha _5=\alpha _8= 0.3246484886897602,   \\ \nonumber
    &&\alpha _6=\alpha _7= 0.1096884778767498.   \nonumber
\end{eqnarray}

Now, we take $h=1$ in our numerical tests. The parameters are given by $E=0.995$, $L=4.6$, $\alpha =0.12$ and $\beta =5.8\times
10^{-4}$. The initial conditions $p_r =0$ and $\theta = \pi /2$.
Given the initial separation $r$, the initial value of $p_\theta$
$(>0)$ is determined in terms of
Equations (9) and (10). For Orbit 1, $r=15$.  For Orbit 2, $r=110$. The two orbits are
integrated by the second-order method $S_2$ and are plotted on the Poincar\'{e} section at the
plane $\theta = \pi /2$ with $p_\theta >0$ in Figure 1a. Clearly, the two orbits have distinct phase
space structures on the Poincar\'{e} section. The structure of Orbit 1 exhibiting a closed curve
describes the regularity of Orbit 1. However, the structure of Orbit 2 consisting of many
points that are randomly distributed in a certain region shows the chaoticity of Orbit 2.
These orbital phase space structures described by the fourth-order methods $S_4$ and $PRK_64$
are almost consistent with those obtained from $S_2$. However, the three integrators have
different orders of magnitude in the Hamiltonian errors $\Delta H=H+1/2$. For the regular
orbit 1 in Figure 1b, the error of $S_4$ is about three orders of magnitude smaller than that
of $S_2$, but about three orders of magnitude larger than that of $PRK_64$. The three methods
are approximately the same in the errors without secular drifts. On the other hand, $S_2$
still shows no secular growth in the error, whereas $S_4$ and $PRK_64$ show secular growths in
the errors for the chaotic orbit 2 in Figure 1c. Such secular drifts are caused by the rapid
accumulation of roundoff errors. Because of this, the errors of $S_4$ and $PRK_64$ approach the
error of $S_2$ in a long enough integration time. $S_2$ is greatly superior to $S_4$ and $PRK_64$ in the
computational efficiency. Considering the computational accuracy and efficiency, we select
$S_2$ as a numerical tool in our later discussions.

\subsection{Orbital Dynamical Behavior}

Figure 2 plots the Poincar\'{e} sections when several different values are given to the
modified gravity parameter $\alpha$. Orbit 2 with the initial separation $r=110$ that is chaotic
in Figure 1a is still chaotic for $\alpha=0.05$ in Figure 2a, $\alpha=0.1$ in Figure 2b and $\alpha=0.18$ in
Figure 2c. We also show a path for Orbit 3 with the initial separation $r=60$ in Figure 1a
going from order to chaos as $\alpha$ increases. This orbit evolves from one single torus for
$\alpha=0.05$ in Figure 2a to six islands for $\alpha=0.1$ in Figure 2b, and to chaos for $\alpha=0.12$
in Figure 1a and $\alpha=0.18$ in Figure 2c. Seen from the global phase space structures in
Figures 1a and 2, an increase of $\alpha$ leads to more orbits with stronger chaoticity. This does
not mean that a given orbit always becomes stronger and stronger chaos in this case.

The regularity of Orbit 1 and the chaoticity of Orbit 2 in Fig.
1(a) can also be identified in terms of the technique of fast
Lyapunov indicator (FLI) in Fig. 3(a). The FLI with two nearby
orbits is defined in [41,42] as a spacetime coordinate independent
indicator
\begin{eqnarray}
\textrm{FLI}=\log_{10}\frac{d(\tau)}{d(0)},
 \end{eqnarray}
where $d(0)$ is the initial proper distance between two nearby
orbits and $d(\tau)$ is a proper distance at the proper time
$\tau$. The FLI of Orbit 1 increasing in a power law with time
$\log_{10}\tau$ shows the regularity of bounded Orbit 1. The FLI
of Orbit 2 increasing in an exponential law with time indicates
the chaoticity of bounded Orbit 2. It is found that ordered orbits
correspond to the FLIs not more than 4.5 and chaotic orbits
correspond to the FLIs larger than 4.5 when the integration time
$\tau=10^6$. With the aid of FLIs, the values of $\alpha$ can be
classified according to the regular and chaotic two cases. In Figure 
3b, the values of $\alpha<0.178$ correspond to order, and the
values of $0.225<\alpha<0.306$ or $\alpha>0.32$ correspond to
chaos.

The method of FLIs is used to find chaos by scanning one parameter
space in Figure 3. This operation is still useful by scanning two
parameter spaces. Figure 4a plots the regions of
$(\alpha,\beta)$ for order and those for chaos. It is shown that
the chaoticity is strengthened when $\alpha$ and $\beta$ increase.
This result is also supported by the method of Poincar\'{e}
sections in Figures 5a-c. Similarly, chaos becomes stronger
with an increase of $E$, as shown in Figures 4b and 5d-f.
However, chaos is weakened when $L$ increases in Figures 4c and
5g-i.

In short, chaos is strengthened from the global phase space
structures when each of the modified gravity parameter $\alpha$,
the magnetic field parameter $\beta$ and the particle energy $E$
increases. However, it is weakened as the particle angular
momentum $L$ increases. Here an explanation is given to these
results. The analysis is based on Eq. (12) with several main terms:
\begin{eqnarray}
H_1 &\approx& -\frac{E^2+L\beta}{2}-\frac{E^2}{r}(1+\alpha)
+\frac{E^2}{2r^2}\alpha(1+\alpha) \nonumber \\
&& +\frac{\beta^2}{8}r^2\sin^2\theta
+\frac{L^2}{2r^2\sin^2\theta}+\cdots.
\end{eqnarray}
The expression is considered when $2(1+\alpha)/r\ll 1$ and
$\alpha(1+\alpha)/r^2\ll 1$. The second term in
Equation (23) corresponding to the second term of Equation (2)
acts as the black hole gravity to the particle. Here, the modified
gravity parameter $\alpha$ also plays a role in enhancing the
gravitational effect. However, $\alpha$ in the third term of
Equation (23) corresponding to the third term of Equation (2)
gives a gravitational repulsive force contribution to the
particle. The gravitational force from $\alpha$  in the second
term of Equation (23)  is more important than the gravitational
repulsive force from $\alpha$  in the third term of Equation (23). 
The fourth term in Equation (23) corresponds to a magnetic field
force as a gravitational effect. The fifth term corresponds to an
inertial centrifugal force from the particle angular momentum.
When anyone of the three parameters $\alpha$, $\beta$ and $E$
increases, the gravitational effects are strengthened and the
particle motions have more dramatic changes. As a result, chaos
would become stronger  from the global phase space structures when
chaos can occur. On the contrary,  an increase of the angular
momentum leads to enlarging the repulsive force effects;
equivalently, it weakens the gravitational effects and decreases
the strength of chaos.

\section{Conclusions}

With the aid of the scalar-tensor-vector modified
gravitational theory, the modified gravity Schwarzschild  black
hole was obtained in the literature through a modified gravity
parameter. This parameter plays important roles in enhancing the
gravitational constant and providing the black hole charge with a 
gravitational repulsive force contribution. The modified
Schwarzschild  black hole is still a static spherically symmetric
black hole solution of the field equation. When the black hole is
immersed in an external asymptotic uniform magnetic field, the
dynamics of charged particles moving in the background field is
not integrable.

Although the Hamiltonian for the description of the charged
particle dynamics is inseparable to the variables, it still allows
for the acceptance of explicit symplectic integrators because the
Hamiltonian  has five splitting parts with analytical solutions as
explicit functions of proper time. Numerical tests show that the
explicit symplectic integrators exhibit good performance in the
long-term conservation of energy integral when appropriate time
steps are chosen.

One of the explicit symplectic integrators combined with the
techniques of Poincar\'{e} sections and fast Lyapunov indicators
is mainly used to survey the effect of the modified gravity
parameter on  the regular and chaotic dynamical features of
charged particle orbits. It is shown that chaos is strengthened
from the global phase space structures under some circumstances as
the modified gravity parameter increases. Such a similar result is
also suitable for the case of the magnetic field parameter and the
particle energy increasing. However, chaos is somewhat weakened
with an increase of the particle angular momentum.

\textbf{Author Contributions}: Software and Writing-original
draft, D.Y.; Software, W.C.;  Formal Analysis, N.Z.;
Investigation, H.Z.; Resources, W.L.; Supervision,
Conceptualization, Methodology, Writing - Review $\&$ Editing and
Funding Acquisition, X. W. All authors have read and agreed to the
published version of the manuscript.

\textbf{Funding}: This research has been supported by the National
Natural Science Foundation of China (Grant Nos. 11973020 and
11533004), and the Natural Science Foundation of Guangxi (Grant
No. 2019JJD110006).

\textbf{Data Availability Statement}: Our paper is a theoretical
work. All of the data are calculated and given in the paper.

\textbf{Institutional Review Board Statement}: Not applicable.

\textbf{Informed Consent Statement}: Not applicable.

\textbf{Acknowledgments}: The authors are very grateful to the
referees for useful suggestions.

\textbf{Conflicts of Interest}: The authors declare no conflict of
interest.

\begin{figure*}
    \centering{
\includegraphics[width=12pc]{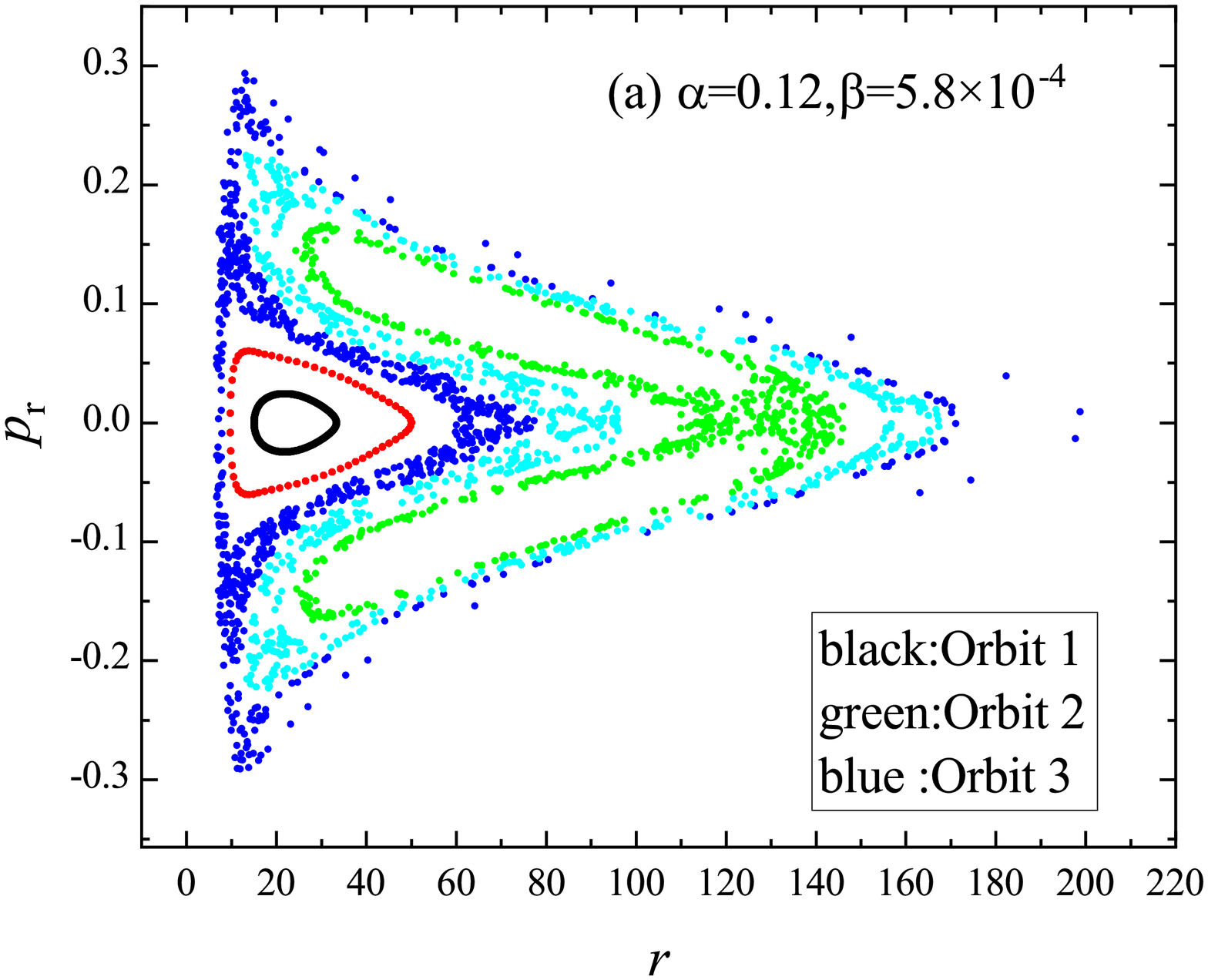}
\includegraphics[width=12pc]{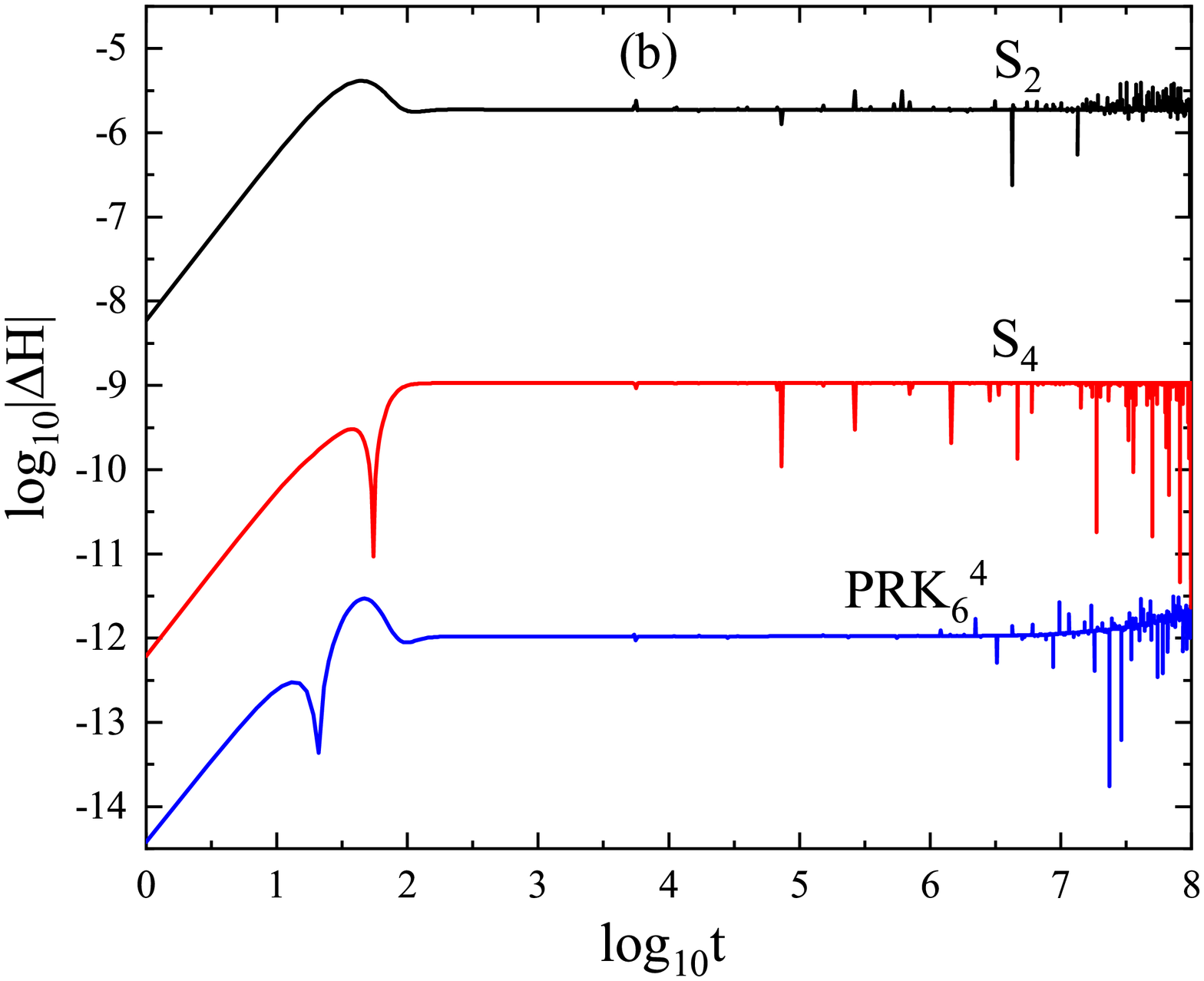}
\includegraphics[width=12pc]{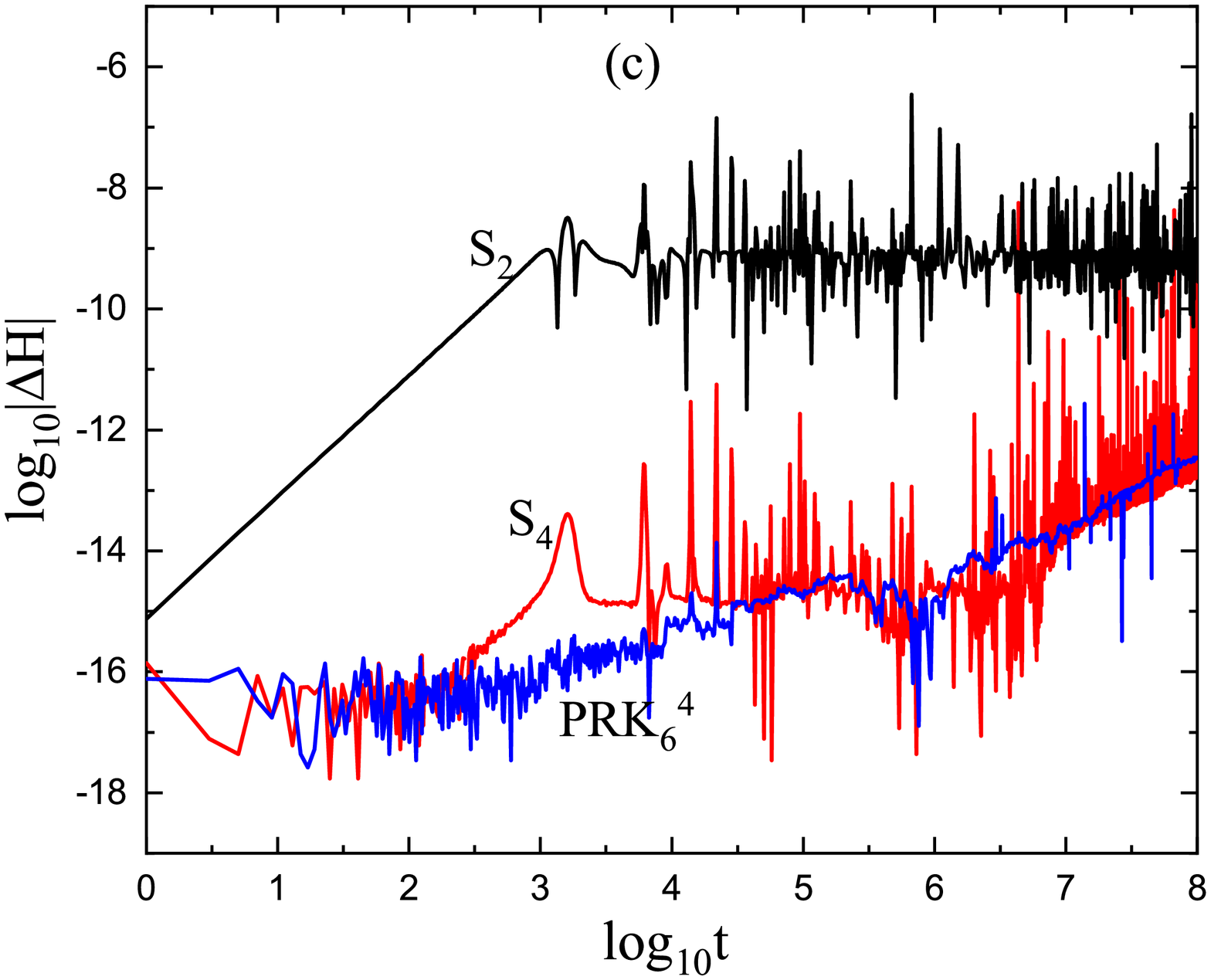}
\caption{(a) Poincar\'{e} sections at the plane $\theta = \pi /2$
with $p_\theta >0$, where the phase space structures are described
by the algorithm $S_2$ with the proper step $h=1$.  The parameters
are $E=0.995$, $L=4.6$, $\alpha =0.12$ and $\beta =5.8\times
10^{-4}$. The initial conditions are $p_r =0$ and $\theta = \pi
/2$. Orbit 1 with the initial separation $r=15$ is a closed
regular curve, and  Orbit 2 with the initial separation $r=110$ is
chaotic. Orbit 3 with the initial separation $r=60$ is also
chaotic. (b) Hamiltonian errors $\Delta H=H+1/2$ for the three
symplectic methods acting on the ordered Orbit 1. The error for
$S_4$ is about three orders of magnitude smaller  than for $S_2$,
but larger than for $PRK_64$. (c) Hamiltonian errors $\Delta
H=H+1/2$ for the three symplectic methods acting on the chaotic
Orbit 2. Although $S_4$ and $PRK_64$ have higher accuracies than
$S_2$ in an integration time, they would approach $S_2$  in
secular error behavior due to roundoff errors.
        }
    }
\end{figure*}

\begin{figure*}
    \centering{
\includegraphics[width=12pc]{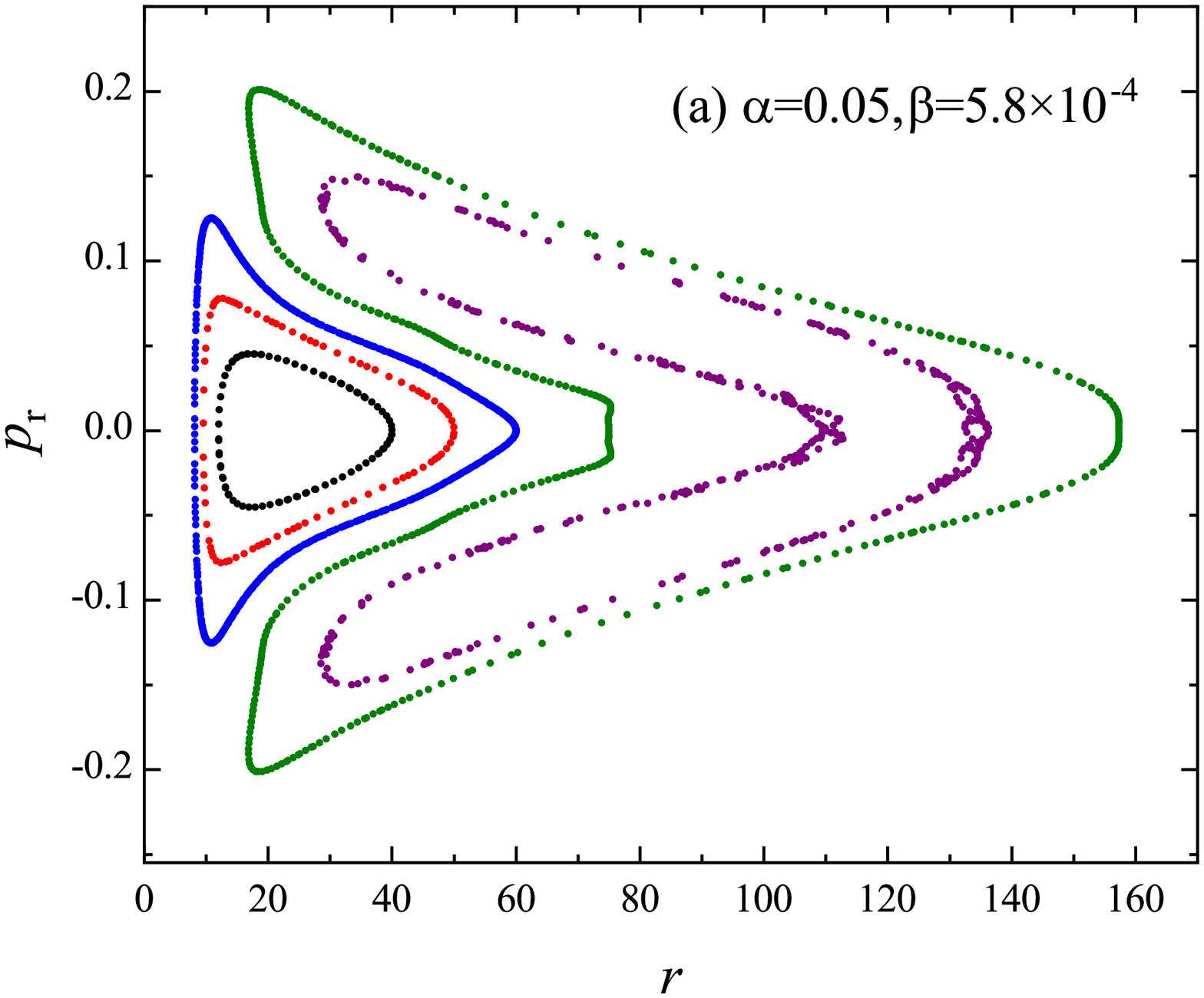}
\includegraphics[width=12pc]{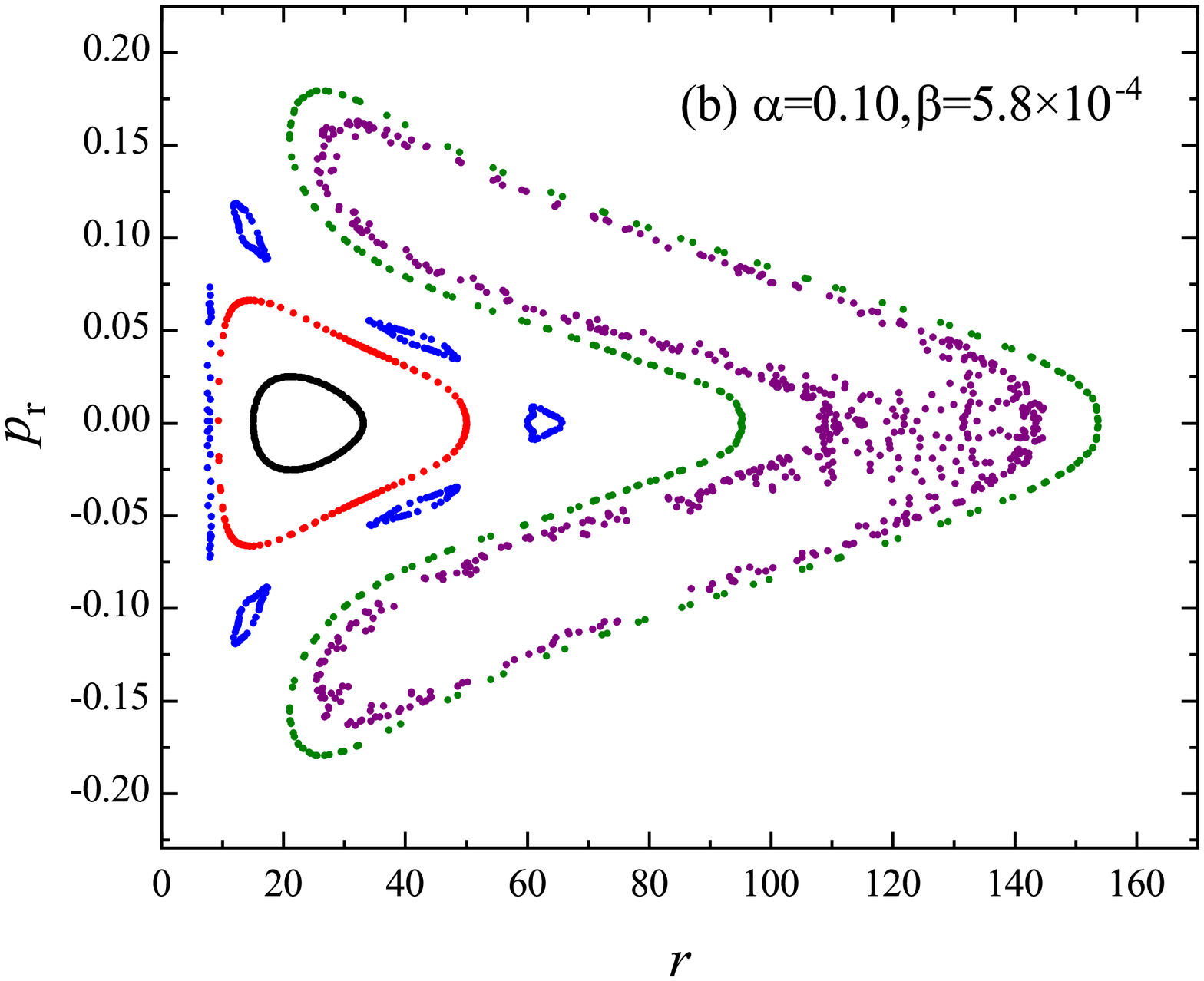}
\includegraphics[width=12pc]{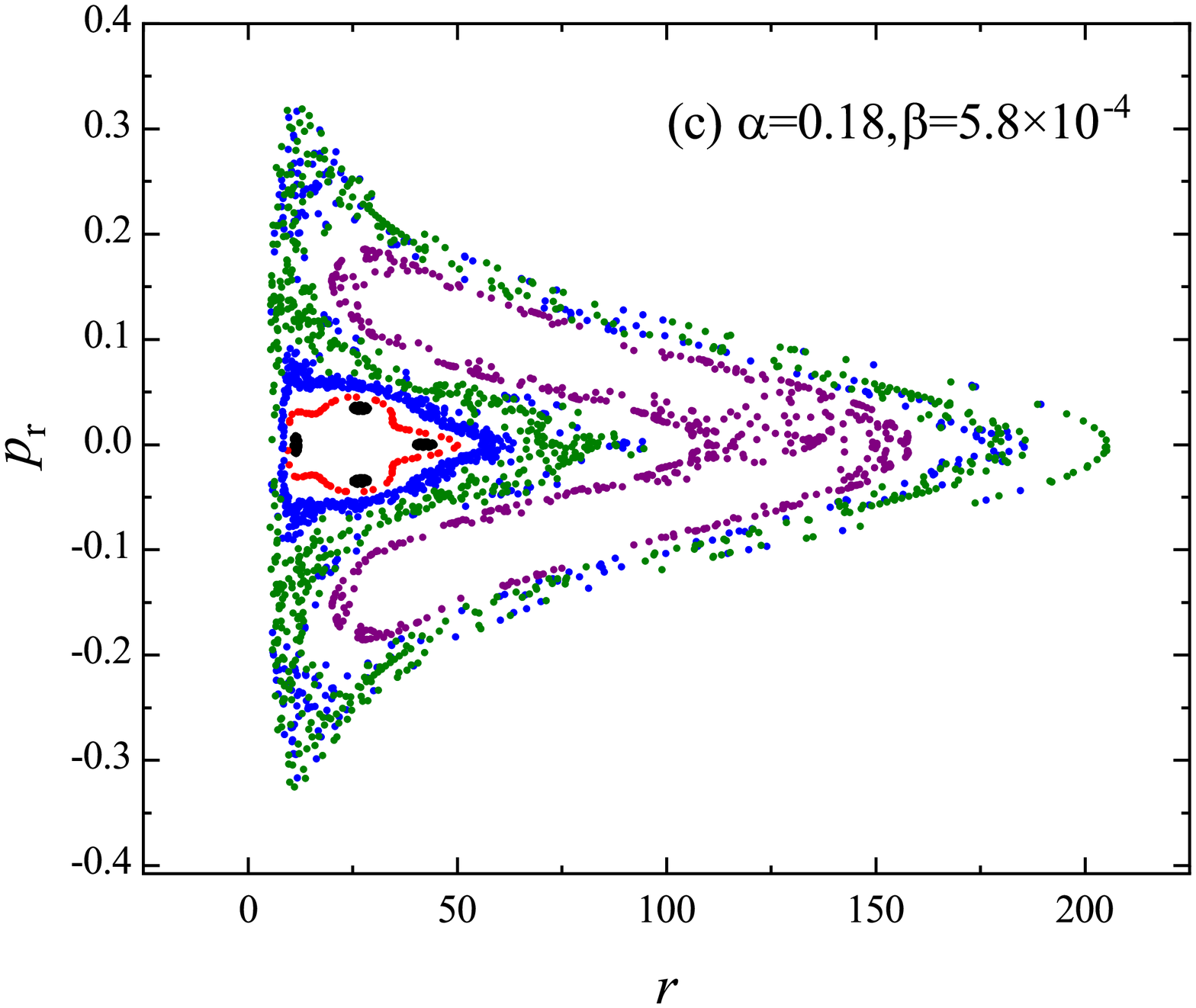}
\caption{Poincar\'{e} sections for three values of the modified
gravity coupling parameter, where $\alpha=0.05$ in (a),
$\alpha=0.1$ in (b) and $\alpha=0.18$ in (c). The other parameters
are $E=0.995$, $L=4.6$ and $\beta =5.8\times 10^{-4}$. It is shown
via the three panels that chaos gets stronger with $\alpha$
increasing.
        }
    }
\end{figure*}

\begin{figure*}
    \centering{
\includegraphics[width=16pc]{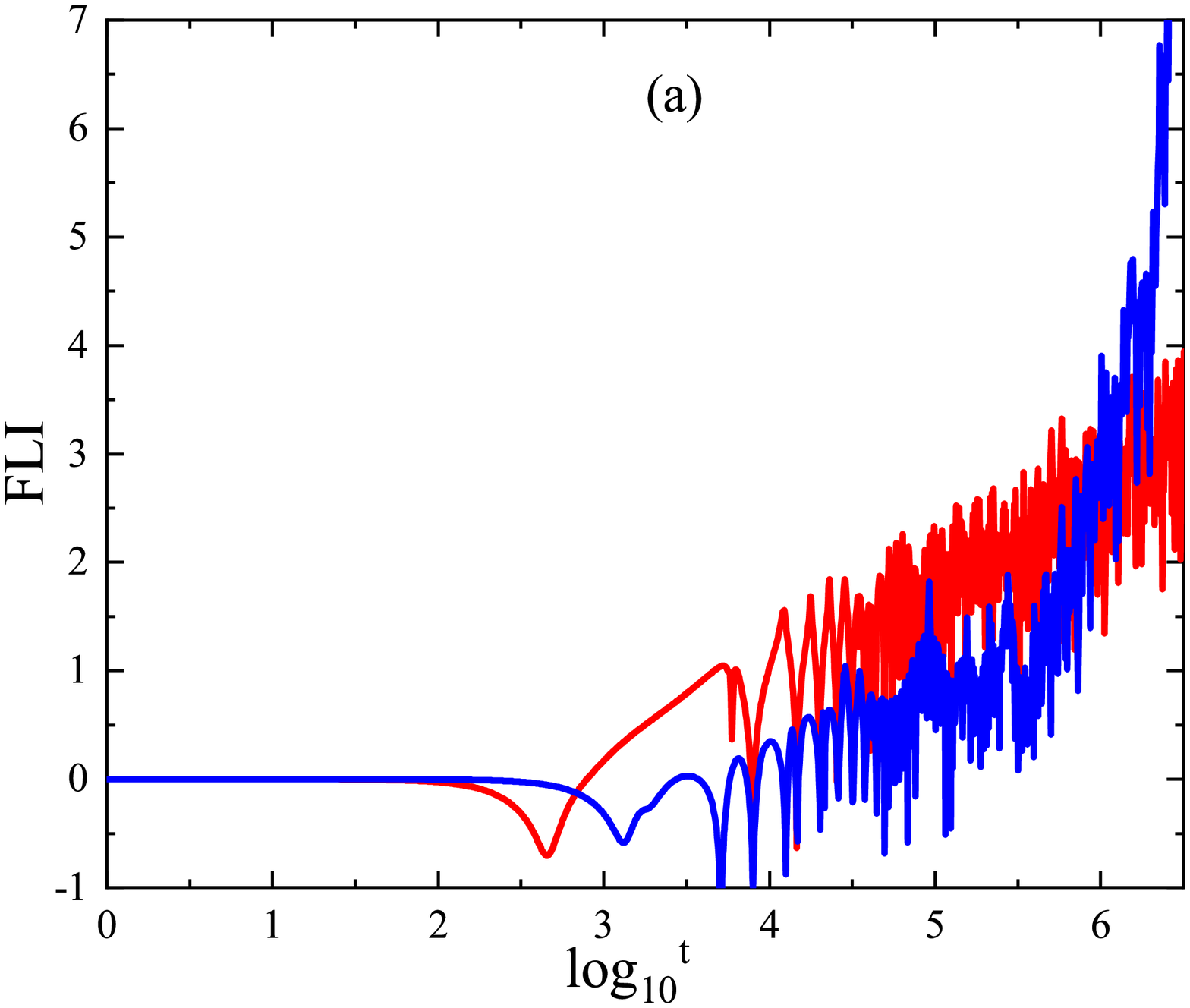}
\includegraphics[width=16pc]{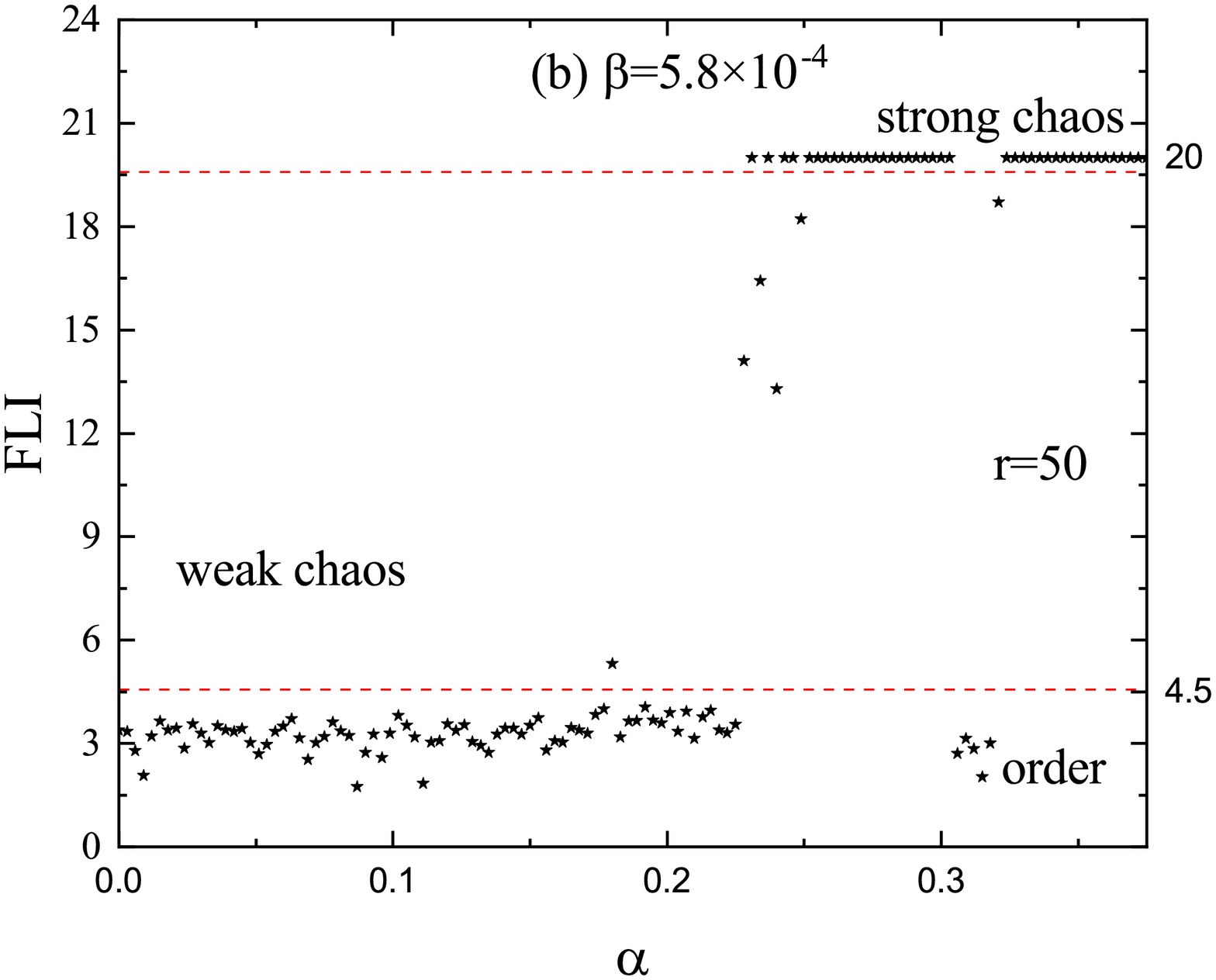}
\caption{(a) Fast Lyapunov indicators (FLIs) for Orbits 1 and 2 in
Fig. 1(a). The FLI grows slowly with time for the ordered orbit 1
(colored Red), but quickly for the chaotic orbit 2 (colored Blue).
(b) Dependence of FLI on the modified gravity parameter $\alpha$.
The other parameters are the same as those of Orbit 1, and the
initial separation is $r=50$. Each value of the FLIs is obtained
after the integration time reaches $\tau=1\times 10^6$. 4.5 is a
threshold value of the FLIs between order and chaos. The FLIs
smaller than (or equal to) this threshold show the regularity.
However, the FLIs more than this threshold indicate the
chaoticity. In this way, the values of $\alpha$ corresponding to
order and those corresponding to chaos are clearly listed in this
figure.
        }
    }
\end{figure*}

\begin{figure*}
    \centering{
\includegraphics[width=12pc]{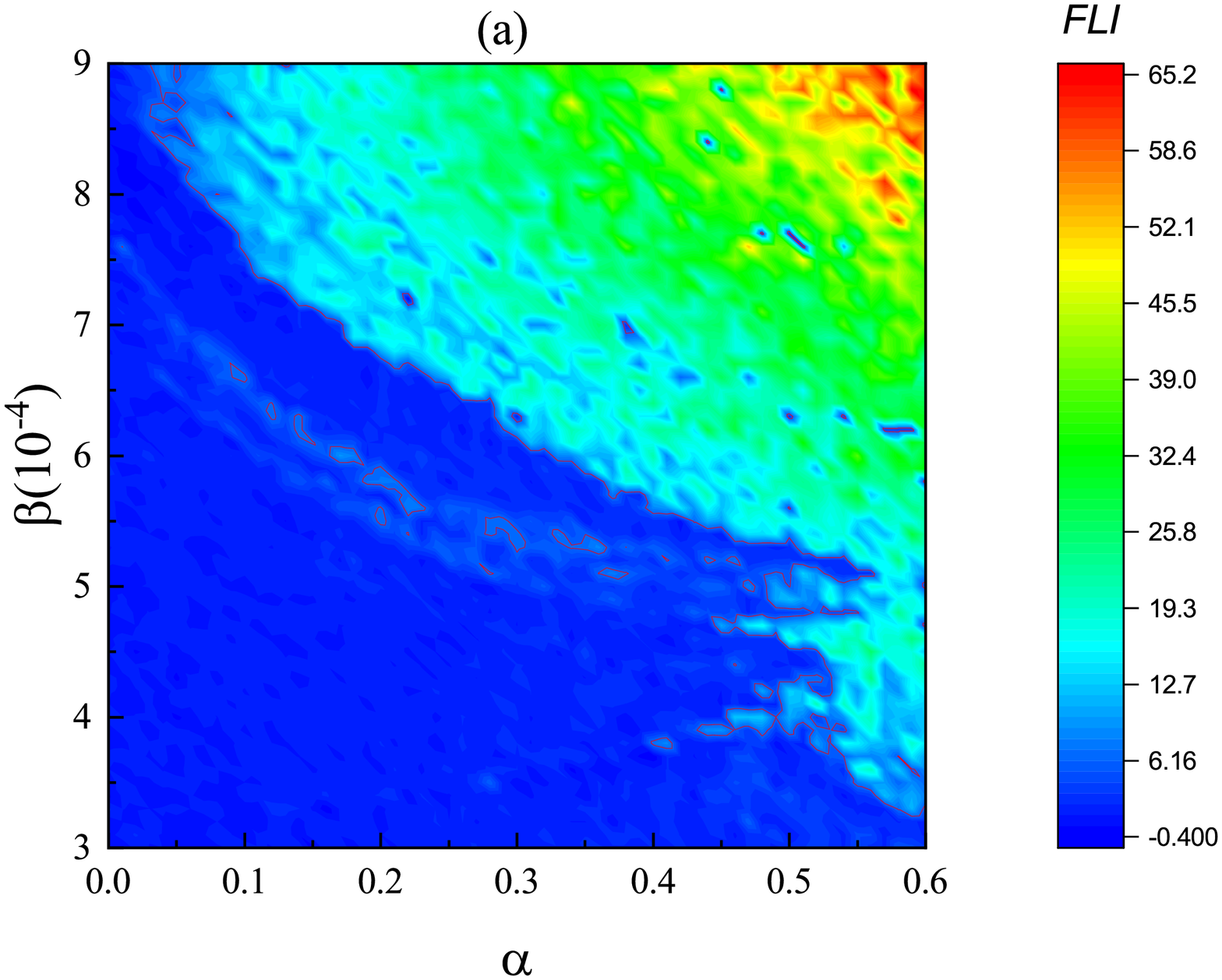}
\includegraphics[width=12pc]{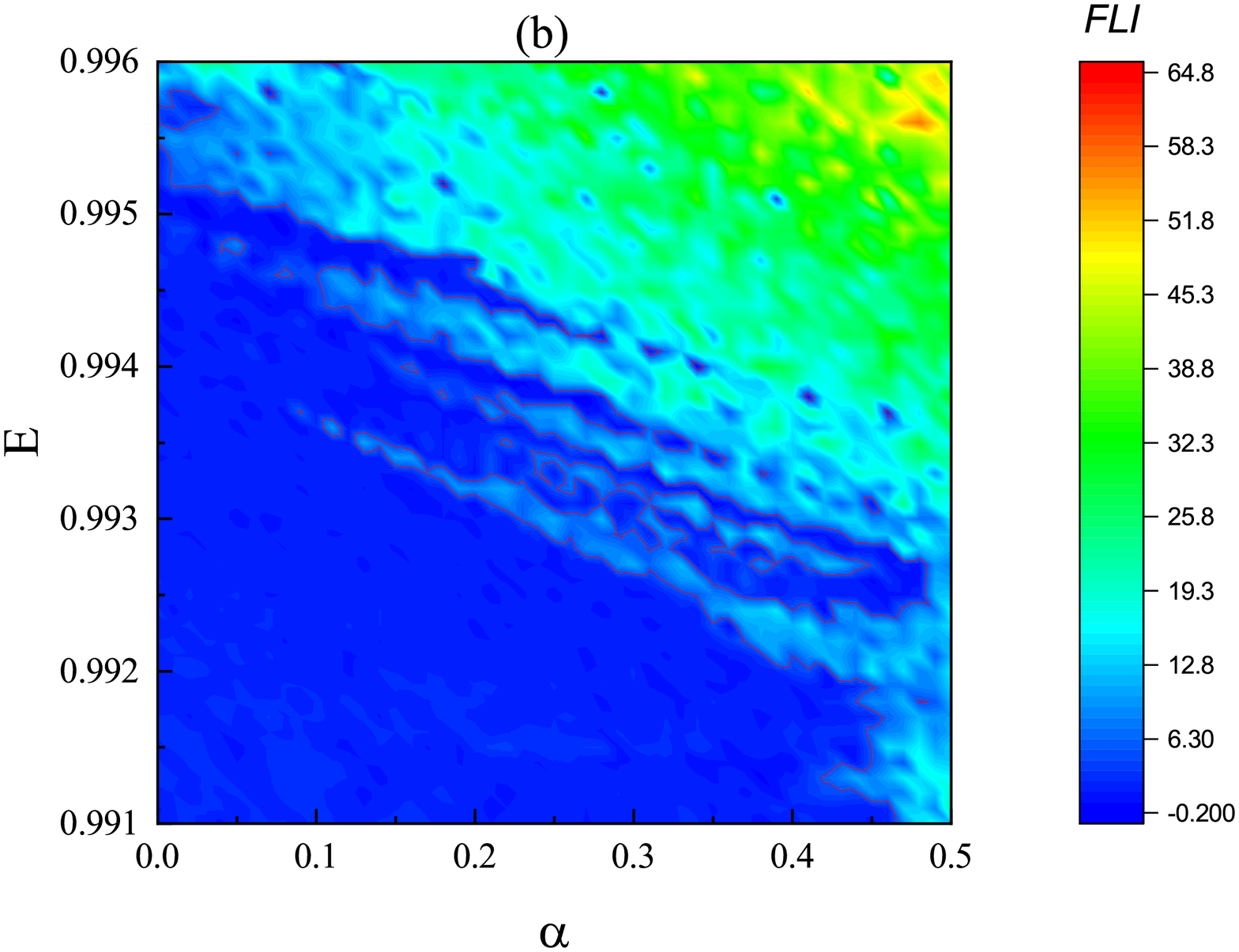}
\includegraphics[width=12pc]{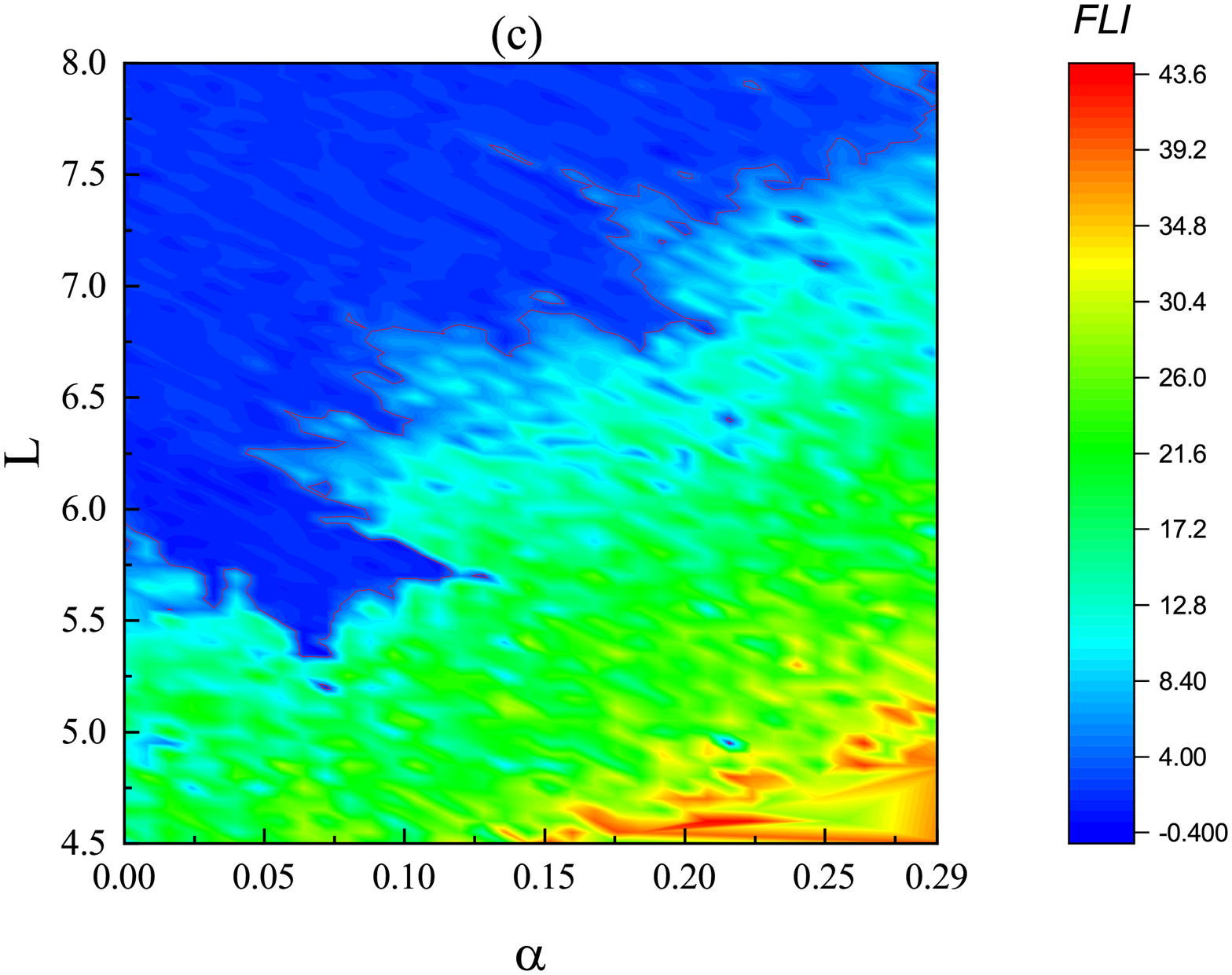}
\caption{Distributions of two parameters corresponding to order
and chaos in terms of FLIs.  The initial radius is $r=110$. (a)
Distributions of $\alpha$ and $\beta$. The other parameters are
$E=0.995$ and $L=6$. (b) Distributions of $\alpha$ and $E$. The
other parameters are $\beta=7.8\times 10^{-4}$ and $L=6$. (c)
Distributions of $\alpha$ and $L$. The other parameters are
$\beta=7.8\times 10^{-4}$ and $E=0.995$. These figures show that
chaos becomes stronger as $\alpha$, $\beta$ and $E$ increase or
$L$ decreases.}
    }
\end{figure*}

\begin{figure*}
    \centering{
        \includegraphics[width=12pc]{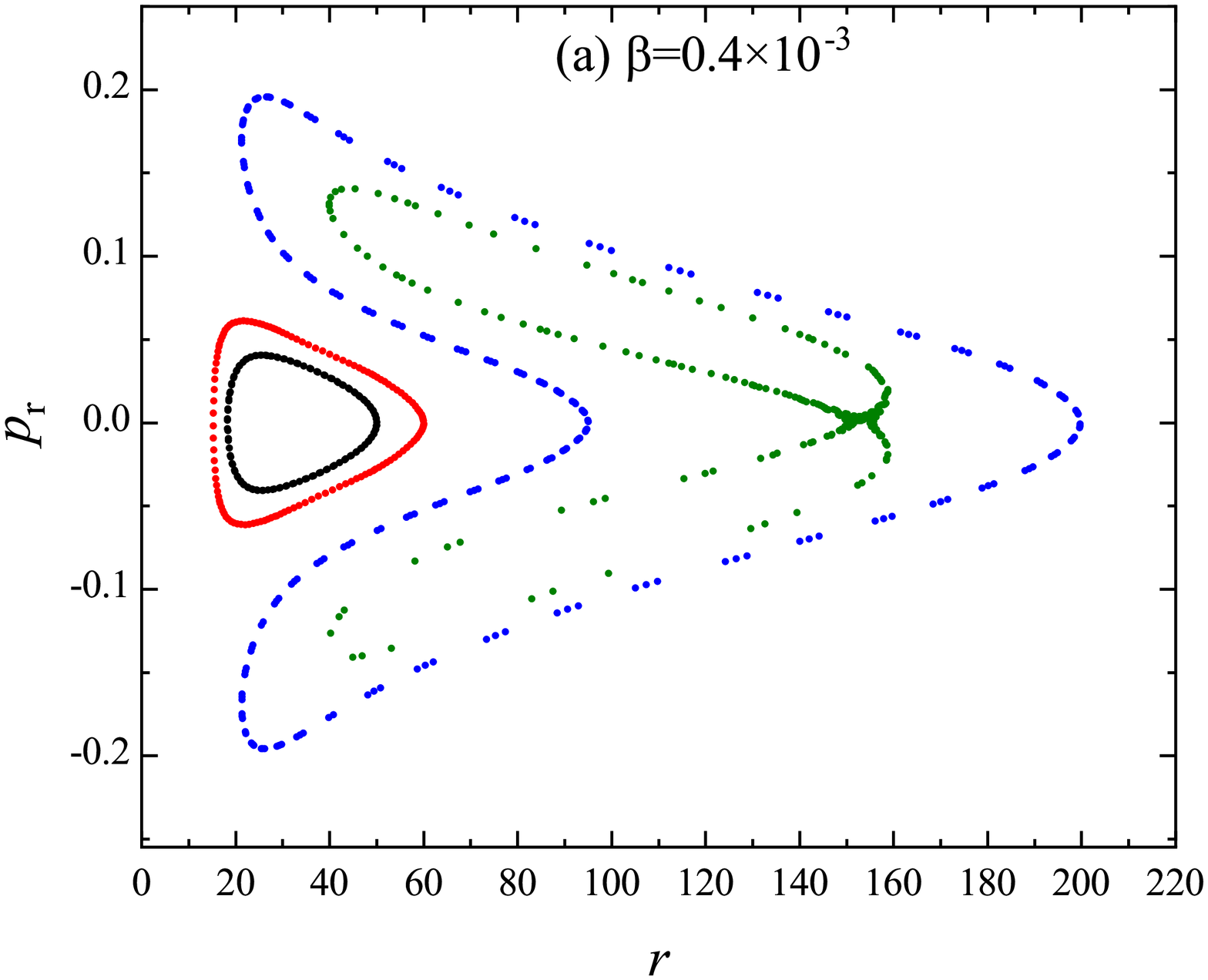}
        \includegraphics[width=12pc]{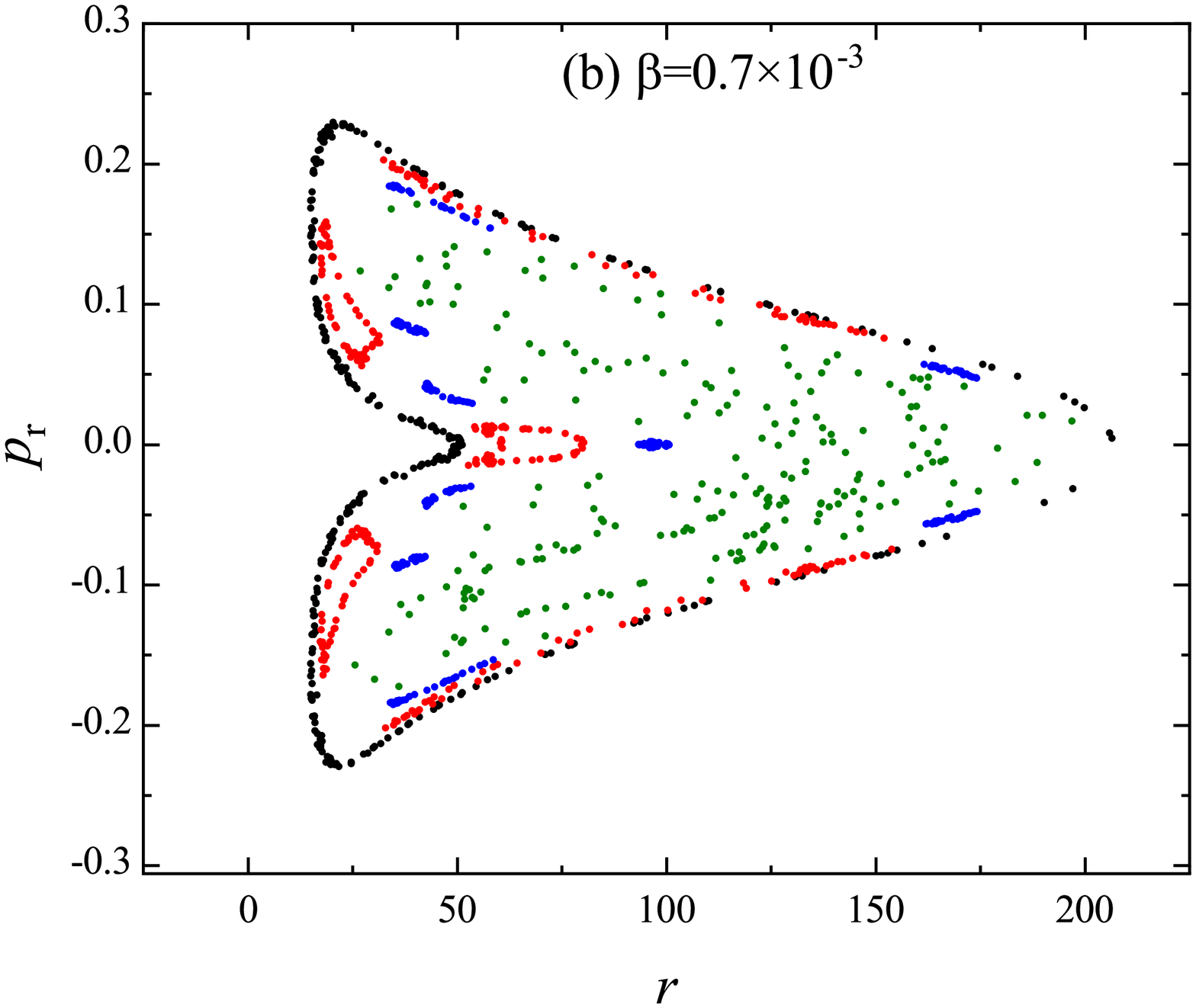}
        \includegraphics[width=12pc]{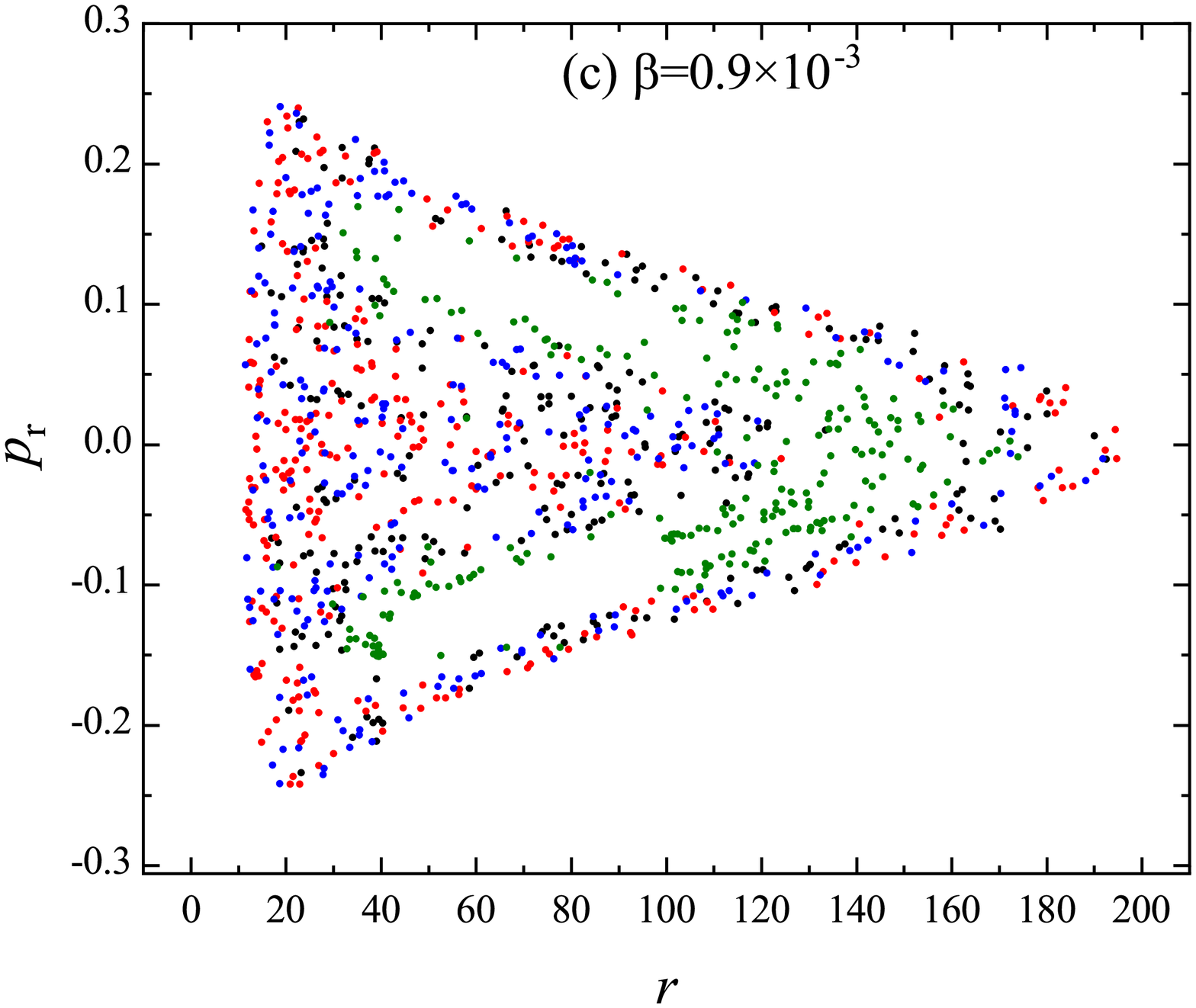}
        \includegraphics[width=12pc]{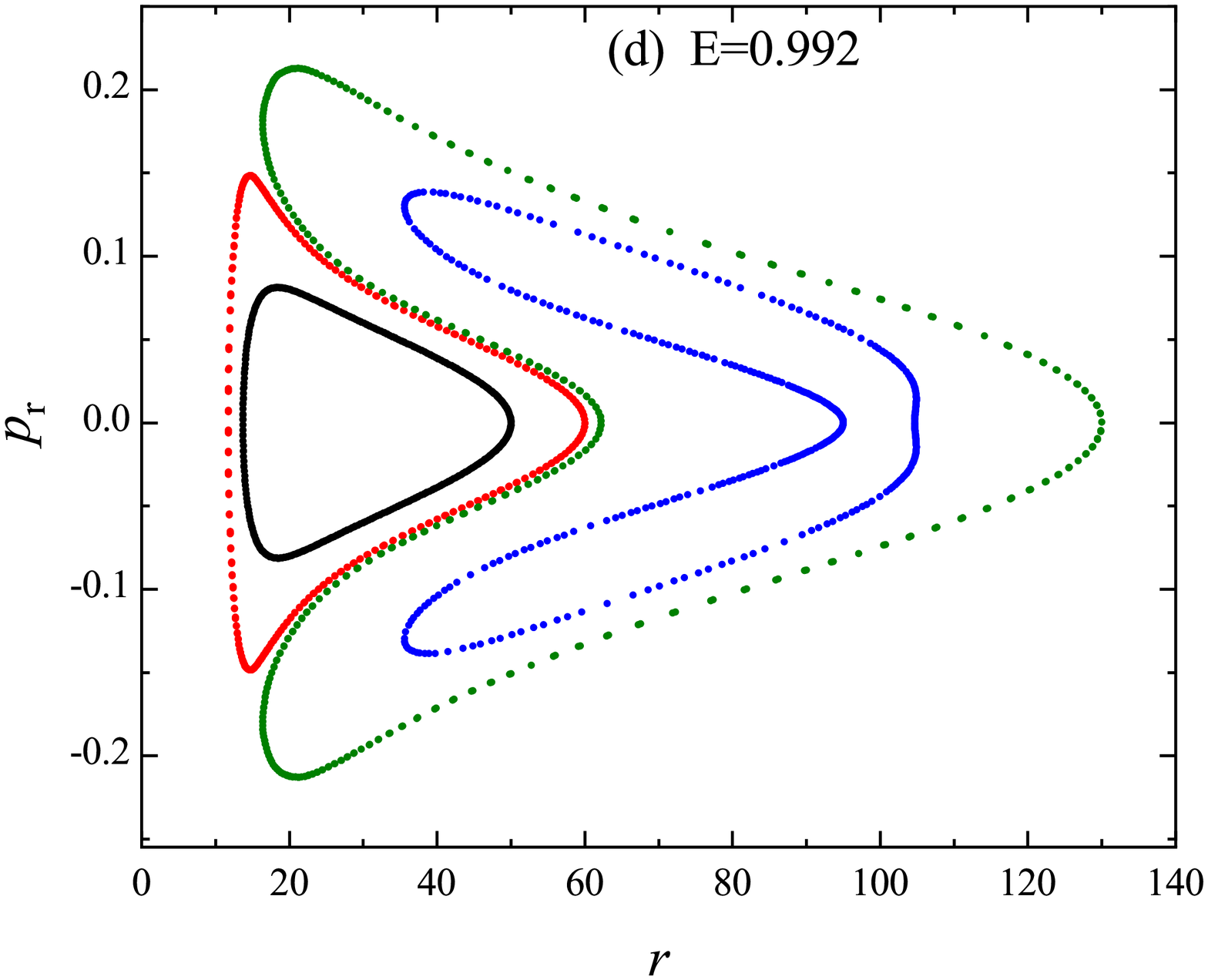}
        \includegraphics[width=12pc]{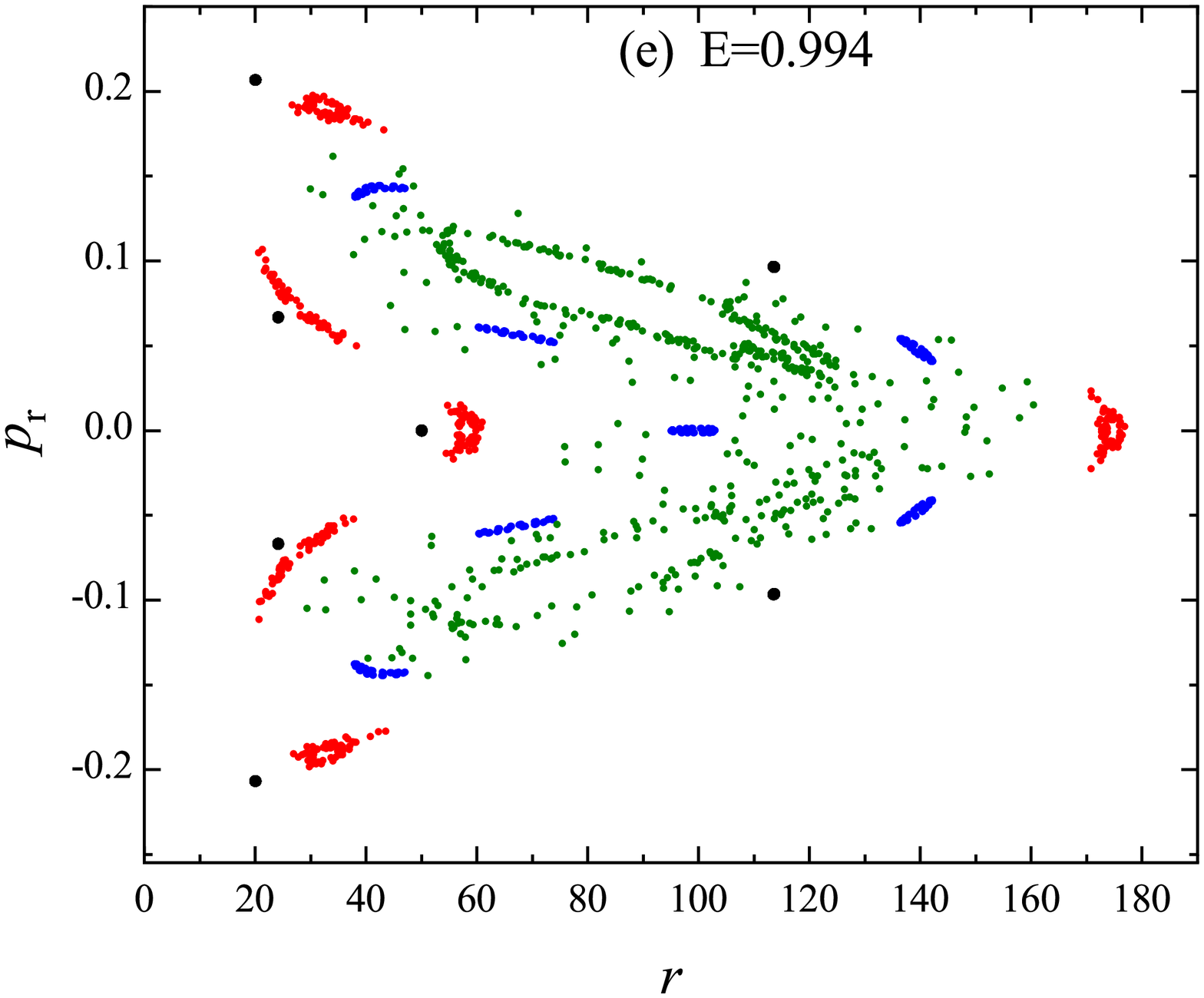}
        \includegraphics[width=12pc]{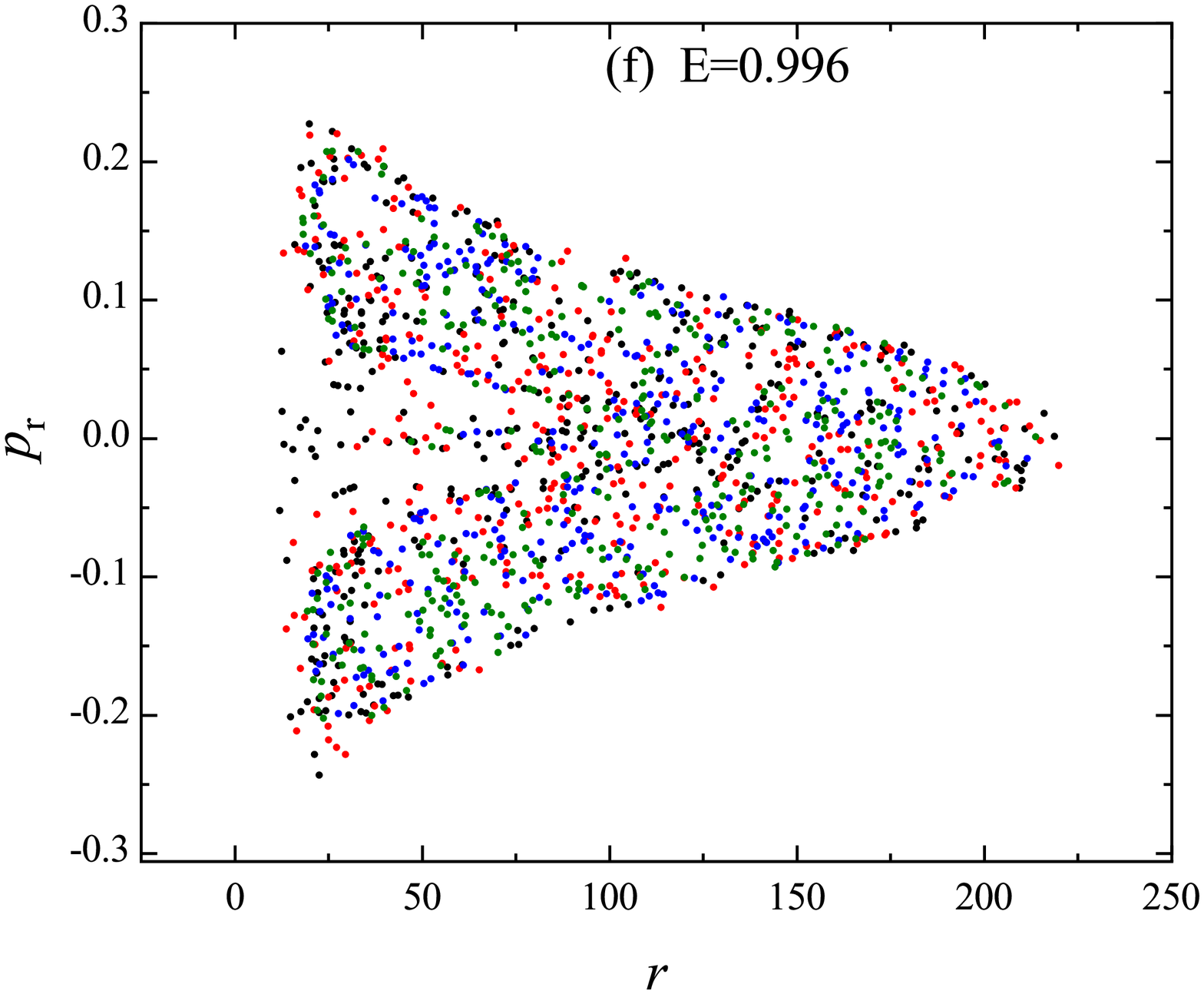}
        \includegraphics[width=12pc]{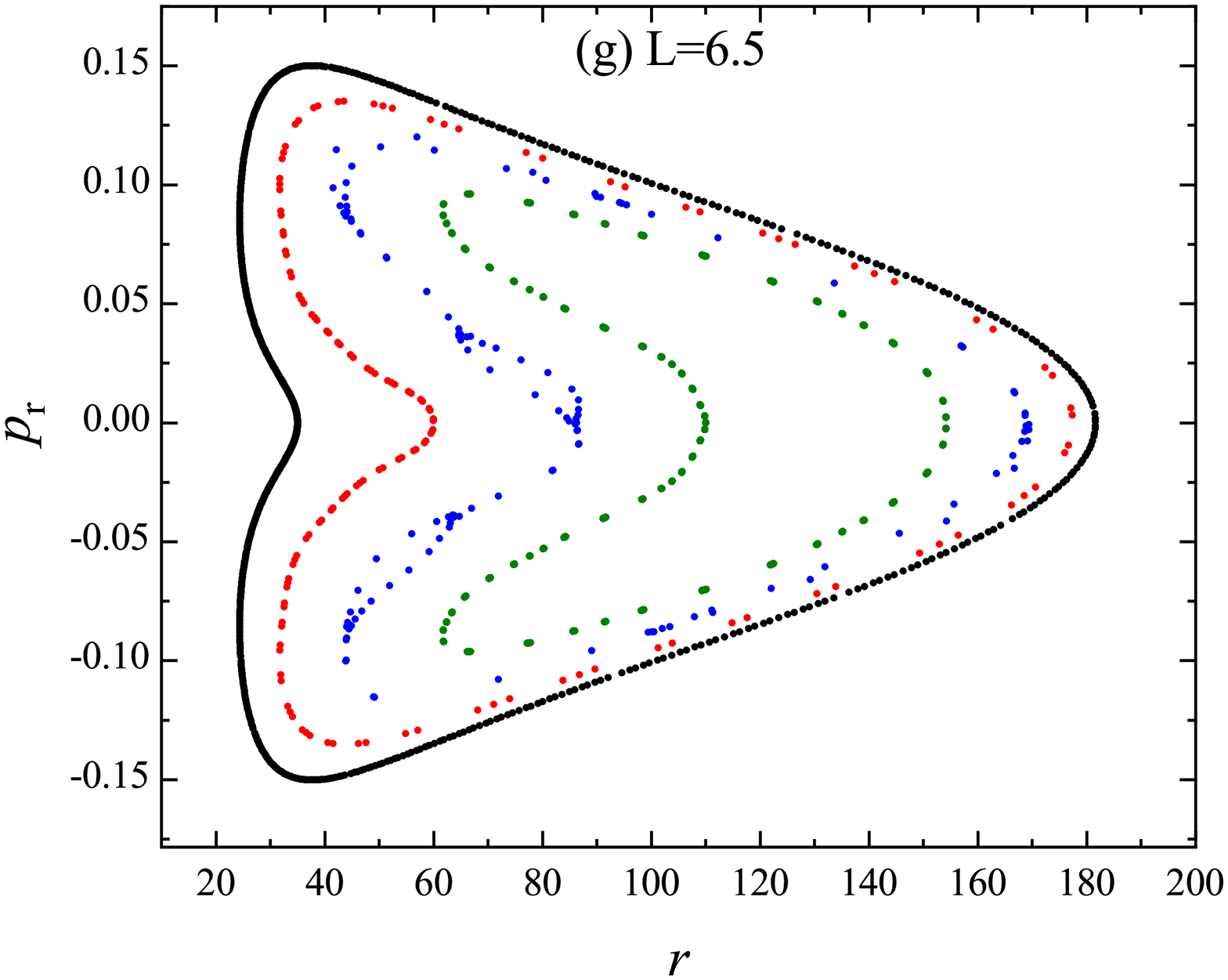}
        \includegraphics[width=12pc]{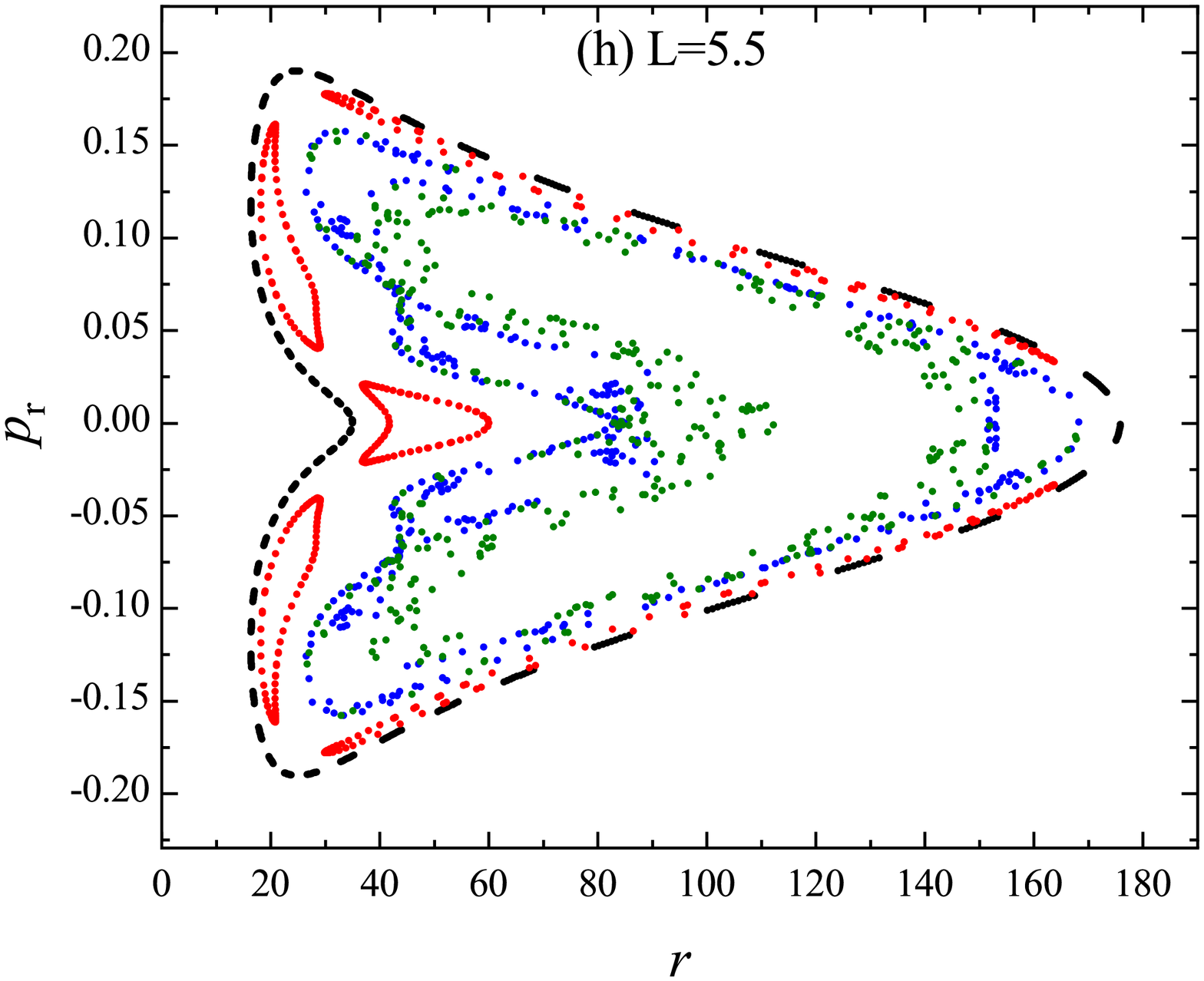}
        \includegraphics[width=12pc]{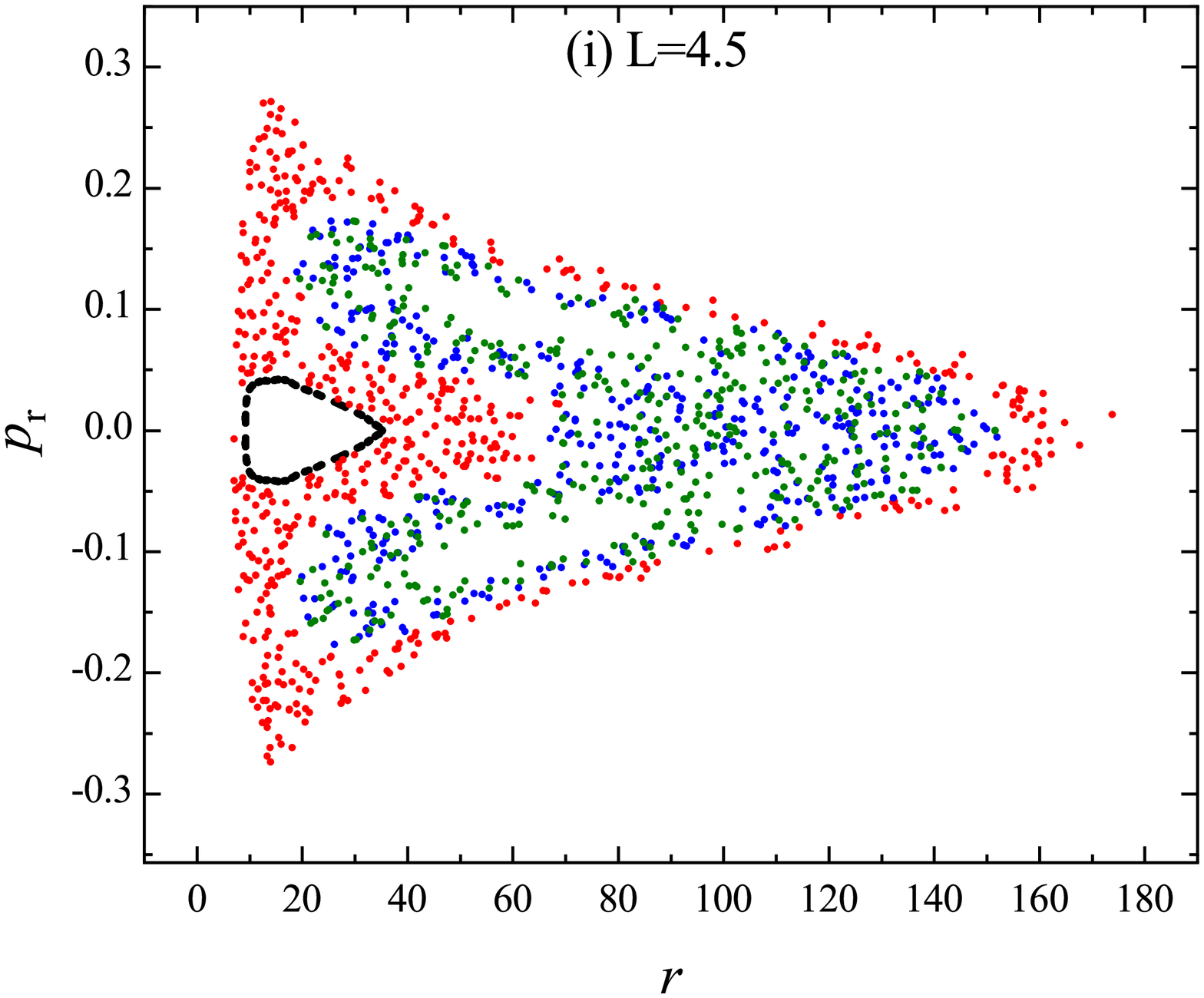}
\caption{Poincar\'{e} sections. (a-c): Three different values
are given to $\beta$. The modified gravity  coupling parameter is
$\alpha =0.3$, and the other parameters are consistent with those
of Fig. 4a. (d-f): Three different values are given to $E$.
The modified gravity  coupling parameter is $\alpha =0.3$, and the
other parameters are those of Fig. 4b. (g-i): Three different
values are given to $L$. The modified gravity  coupling parameter
is $\alpha =0.05$, and the other parameters are those of Fig.
4c.
        }
    }
\end{figure*}

\end{document}